\documentclass[preprint,12pt, a4paper]{elsarticle}

\usepackage{amssymb}
\usepackage{hyperref}

\usepackage[dvipsnames]{xcolor}

\definecolor{codebg}{RGB}{240,240,240}
\usepackage{listings}
\lstdefinestyle{listingstyle}{
    breakatwhitespace=false, 
    tabsize=2,
    keywordstyle=\bfseries\color{black},
    commentstyle=\itshape\color{gray},
    basewidth = {.47em},
    stringstyle=\color{Sepia},
    basicstyle=\small\ttfamily\color{darkgray},
    showstringspaces=false
}
\usepackage{tabularx}
\newcommand*\rot{\rotatebox{78}}
\newcommand*\cm{\checkmark}

\usepackage{xspace}
\usepackage{relsize}
\newcommand{\Rplus}{\protect\hspace{-.05em}\protect\raisebox{.30ex}{\smaller{\smaller\textbf{+}}}}
\newcommand{\cpp}{\mbox{C\Rplus\Rplus}\xspace}

\journal{SoftwareX}

\begin{document}

\begin{frontmatter}

\title{Reticula: A temporal network and hypergraph analysis software package}

\author{Arash Badie-Modiri\corref{cor1}}
\ead{arash.badiemodiri@aalto.fi}
\ead[url]{https://arash.network}
\cortext[cor1]{Corresponding author}
\author{Mikko Kivelä}

\address{Department of Computer Science, School of Science, Aalto University, FI-0007, Finland}

\begin{abstract}
In the last decade, temporal networks and static and temporal hypergraphs have enabled modelling connectivity and spreading processes in a wide array of real-world complex systems such as economic transactions, information spreading, brain activity and disease spreading. In this manuscript, we present the Reticula C++ library and Python package: A comprehensive suite of tools for working with real-world and synthetic static and temporal networks and hypergraphs. This includes various methods of creating synthetic networks and randomised null models based on real-world data, calculating reachability and simulating compartmental models on networks. The library is designed principally on an extensible, cache-friendly representation of networks, with an aim of easing multi-thread use in the high-performance computing environment.
\end{abstract}


\end{frontmatter}


\section{Motivation}

A wide variety of interesting physical systems consist of a large number of interacting entities with different levels of internal complexity. For example, a social network consists of individuals exchanging information with each other physically or electronically \cite{scott1988trend}. Similarly, a biological system such as a human brain can be viewed as the aggregate of its cells and their connections \cite{rubinov2010complex,fornito2016fundamentals}. Many emergent behaviours of these complex systems can be reproduced and analysed by modelling the system as a graph. This gave rise to the field of complex networks, where this approach is being applied to increase our understanding of phenomena on a vast array of interconnected systems. For example, it has been noted that many systems show patterns of synchronisation between vertices \cite{wiley2006size,mirollo1990synchronization} and that spreading processes in completely different contexts, such as information or disease spreading in a population \cite{davis2020phase,karsai2011small} or accessibility of a public transport system \cite{badie2018error} can show phase-transitions that have been the subject of extensive previous studies in statistical physics.

This increased interest in the study and modelling of real-world systems as static networks with dyadic relations between vertices (graphs) spurred on a set of software libraries aiming to fill the computational gap, such as NetworkX \cite{hagberg2008exploring}, igraph \cite{csardi2006igraph}, and Stanford Network Analysis Project (SNAP) \cite{leskovec2016snap}. These libraries have struck different compromises between API usability, performance and generality.

Many real-world systems, however, consist of connections and entities that evolve through time periods shorter than or comparable to the period of observation. Social circles evolve through time with new connections forming between people. Cells in a biological system die or change their nature and strength of connections with other cells. Furthermore, time dictates an inherent direction to connectivity: unlike a graph, interactions between entities are only transitive if they happen in causally plausible times. Temporal networks \cite{holme2015modern,holme2019temporal} aim to provide a representation of complex systems that include interaction times between the entities, as opposed to what a static network edge usually represents, i.e., the possibility or plausibility of interaction between entities. This set of timestamped interactions (also known as temporal edges or events) inherently captures the evolution of the system through time and all the possible temporal inhomogeneities and correlations that affect the dynamical processes but might be overlooked in static settings \cite{karsai2011small}.

This popularity and conceptual simplicity of temporal networks as a model have brought about another set of software libraries, such as Raphtory, a distributed temporal network analysis library \cite{steer2020raphtory}, PathPy focusing more on temporal paths and path statistics \cite{scholtes2017network} and Phasik, with a strong focus on inferring temporal phases in the network \cite{lucas2021inferring}.

Many systems, on the other hand, are best described using interactions or events that are not constrained to two entities. For example, face-to-face or online interactions in a population regularly happen in groups of size larger than two, and the interaction between scientific literature in form of citation is in fact an interaction between the group of authors of the cited paper and the group of authors in the citing. Specialised libraries have emerged for handling hypergraphs, for example, SimpleHypergraphs.jl \cite{antelmi2019simplehypergraphs} and HyperGraphs.jl \cite{diaz2022hypergraphs} for Julia, and Python packages HyperNetX \cite{joslyn2020hypernetwork} and Complex Group Interactions (XGI) \cite{landry2022xgi}, aimed at analysing higher-order interactions.

Reticula is a \cpp library and the accompanying Python bindings that natively supports a wide range of complex network types, including directed and undirected variations of static and temporal networks, hypergraphs and hypergraph temporal networks. Additionally, it supports delayed events for temporal networks and temporal hypergraphs, allowing users to model, e.g., transportation networks or other temporal networks that involve latency. The library also supports various methods of synthetically generating and randomising different types of networks, reading and writing networks to files, calculating various properties of networks, and inter-operation with NetworkX for static dyadic network types.

\section{Impact}
The Reticula Python interface enjoys a sizeable speedup compared to the pure Python libraries. For example, generating a single random graph with degree sequence $(4, \ldots 4)$ of size 1000 using the \lstinline[language=Python]{random_degree_sequence_graph} function is about 18 times faster in Reticula than NetworkX. Generating 1000 such graphs, while allowing the use of 24 CPU cores in parallel increases this ratio to 250 times. Similarly, Generating a single random expected degree-sequence hypergraph of size 16000, expected vertex degree sequence $(6, \ldots 6)$ and expected edge degree sequence $(3, \ldots 3)$ is about 3.8 times faster in Reticula compared to XGI.

Earlier versions of this software have enabled various works of research. Ref.~\cite{badie2022directedprr} used it to estimate constrained reachability on large real-world temporal networks, while Ref.~\cite{badie2022directed} applied the same methodology to a variety of random temporal network models to show that limited waiting-time adjacency has a directed percolation reachability phase transition in many temporal networks. Furthermore, Ref.~\cite{rizi2022epidemic} utilised the library in its implementation of compartmental model dynamics on networks to predict the effects of degree heterogeneity, homophily and the extent of adaptation of contact tracing apps on their efficacy as a preventative measure.

\section{Design goal and description of the software}
The software package interface is designed to optimise for the following goals: (1) providing an easy to use, human-readable function interface, generally akin to that of NetworkX, that can be uniformly used for all supported network types, and (2) making it easy to use effectively in high-performance computation environment by safe multi-threading, especially in Python, (3) optimising the CPU cache hit rate in the implemented algorithms as much as possible.

While from a theoretical complexity perspective, the commonly used hashtables-nested-in-hashtable representation of graphs offers constant-time complexity for checking the existence of edges between two vertices, this approach results in sub-optimal memory access patterns in many graph algorithms that require looping through all edges incident to a certain vertex. This representation is also not directly extensible to temporal networks and hyper-graphs.

Instead, Reticula stores the set of incident edges for each network vertex in a contiguous area of the memory as a sorted vector. The contiguous storage of incident edges allows the processor to optimally utilise its multi-layer cache system, avoiding slow and costly access of the main system memory. This same structure can be readily expanded to hypergraphs without modifications. Furthermore, it can support directed networks by storing separate in- and out-incident vectors of edges for each vertex. The support for temporal networks and hypergraph temporal network can be included by sorting the incident vector of temporal edges by time.

\begin{table}[h!]
\caption{\label{tab:compare} Comparison of Reticula to other network libraries mentioned in this manuscript. Implementation refers to the main language of implementation of each library. Multi-threading in this table refers to either providing a built-in shared memory parallel processing model (e.g., through OpenMP) or enabling the user to do so by releasing the global interpreter lock and providing thread-safe types and functions. In the Implementation and Additional interfaces rows, ``Py'' refers to the Python programming language.}
\centering
\begin{tabularx}{\linewidth}{lcccccccc}
                        & \rot{Reticula} & \rot{NetworkX}
                        & \rot{SNAP}      & \rot{igraph}
                        & \rot{PathPy}   & \rot{Phasik}
                        & \rot{HyperNetX} & \rot{XGI} \\
\hline\hline \\
Static Networks         & \cm & \cm & \cm & \cm & \cm & -   & -   & -   \\
Temporal Networks       & \cm & -   & -   & -   & \cm & \cm & -   & -   \\
Hypergraphs             & \cm & -   & -   & -   & -   & -   & \cm & \cm \\
Temporal Hypergraphs    & \cm & -   & -   & -   & -   & -   & -   & -   \\
Parallel edges          & -   & \cm & \cm & \cm & -   & -   & \cm & \cm \\
Immutable graph type    & \cm & \cm & \cm & -   & -   & -   & \cm & -   \\
Implementation          & \cpp & Py  & C   & C   & Py  & Py  & Py  & Py  \\
Multi-threading         & \cm & -   & \cm & \cm & -   & -   & -   & -   \\
Additional interfaces   & Py  & -   & Py  & Py+R & -  & -   & -   & -   \\
\end{tabularx}
\end{table}

To make it easier to use this library in a heavily multi-threaded environment without the necessity to use explicit synchronisation primitive the library mostly presents immutable edge and network types, provided alongside a variety of functions for manipulating networks that return a modified copy instead of mutating the object itself. This removes the risk of, e.g., inadvertently modifying a network that is being simultaneously read in another thread which might return incorrect results with no visible errors or even result in undefined behaviour. This, combined with the fact that the \cpp library in most parts does not directly manipulate any Python objects, allows us to safely release the Global Interpreter Lock (GIL) at the entry point of the Python interface, allowing multi-threaded computation directly from Python code.

The library currently supports the representation of various types of static and temporal networks, with directed or undirected connections, either limited to dyadic connections or hyperedges involving any number of vertices. Moreover, the library allows the users to construct higher-order networks, where vertices are themselves edges of a simpler network, to an arbitrary degree for the \cpp library and the second order for the Python library. The Python binding additionally supports integral, string and 2-tuple vertices, while in \cpp the templated types allow any type of vertices that defines strong ordering and certain utility functions for that type, which by default includes all numeric types, strings and all ordered containers. Novel types of networks can easily be implemented in \cpp by defining a custom edge type and their respective type traits.

The \cpp library can be directly compiled and installed on the target system, or preferably directly included in a project using the \verb|FetchContent| CMake module. The Python binding, implemented using PyBind11 \cite{pybind11}, can be installed from the Python Package Index using the console command \lstinline[language=bash]{python -m pip install -U reticula} on any 64-bit Linux operating system with GNU C Library (glibc) version 2.17 or newer (i.e., compatible with the platform tag \verb|manylinux2014|) and Python version 3.8 or newer. Both libraries provide a similar interface and set of types. Although the \cpp library provides requirements and concepts that make it easy for researchers to implement custom functionality, such as new types of edges or networks that can benefit from the already implemented algorithms, the pre-compilation requirements of Python native extension interface makes such on-the-fly extensibility unfeasible.

The core library is extensively and automatically tested. The tests are implemented in the \verb|src/tests/reticula/| directory of the \cpp library source tree. The code-base makes heavy use of many modern \cpp features such as concepts and ranges, which, alongside other best practices such as strict const-correctness provide a level of guarantee against some inadvertent common error-prone patterns by library authors or the end-users, but requires a recent compiler suit with decent support for a subset of the \cpp20 standard (ISO/IEC 14882:2020 \cite{iso2020cpp}) minus the co-routines, modules and the string formatting sections of the standard. For the current release, the library is tested to compile on GNU Compiler Collection versions 10.2 and newer. Extensive documentation for installation and use of the software is available online at \url{https://docs.reticula.network/} and in the \verb|docs/| directory of the Python binding source tree.

\section{Implemented functionality}
The library allows the input and output of networks from edge- or event-list files. It is also possible to import and export static dyadic networks to and from NetworkX. This can be used, for example, to create static networks through methods not currently implemented by this library or read from or write to other file formats.

Reticula can also be used for generating various synthetic and random static and temporal networks, such as regular ring lattices, $d$-dimensional square lattices, $G(n, p)$ \cite{batagelj2005efficient}, Barabási--Albert \cite{barabasi1999emergence}, $k$-regular, (directed or undirected) degree-sequence \cite{bayati2010sequential} and (directed or undirected) expected degree-sequence random graphs \cite{chung2002connected,miller2011efficient}, as well as fully-mixed and activation model temporal networks with any static ``base'' and exponential, geometric, self-exciting and power-law inter-event time distributions \cite{badie2022directed}. It is also possible to start from a (possibly real-world) temporal network and randomise certain features away using various microcanonical randomised reference models implemented in the library \cite{gauvin2022randomized}.

While the library primarily focuses on spreading, connectivity and reachability analysis of networks, it implements various other well-known network algorithms that might be of use as building blocks of other algorithms or measurements on the network. For example, the library allows users to check whether a degree (pair) sequence is (di-)graphical \cite{erdos1960graphs,kleitman1973algorithms}, and whether a directed (hyper-)graph is acyclic and to find a topological ordering of directed acyclic graphs. The user can calculate the density of directed or undirected dyadic static networks. For temporal networks, the user can construct the static projection of the network: a directed or undirected static network where two vertices are connected if they have at least one event in the temporal network. They can also calculate the timeline of all temporal events that correspond to each static projection link.

On the subject of reachability and connectivity, the library provides extensive functionality. For static networks, the user can compute (weak) connectivity and (weakly-) connected components, query whether a vertex is reachable from another and calculate all in- and out-components and shortest-path lengths to and from any vertex. It is also possible to estimate, using probabilistic counting methods, the in- or out-component sizes of a directed network in a single pass in $O(|\mathcal{E}| \log |\mathcal{E}|)$ time for the common case of acyclic graphs. \cite{badie2020efficient}.

Similarly, for temporal networks it is possible to calculate the event graph representation \cite{kivela2018mapping}, and to calculate the temporal reachability cluster starting from a single vertex at a specified time in $O(|\mathcal{E}|)$, or estimate it in one pass for all vertices at all possible starting times in $O(|\mathcal{E}| \log |\mathcal{E}|)$ \cite{badie2020efficient}. The resulting temporal clusters include reachability information, and characteristic quantities such as cluster mass, volume and lifetime \cite{badie2022directed,badie2022directedprr}.

The temporal network reachability cluster calculation can be used with various definitions of temporal adjacency with many of the most commonly used definitions currently implemented. The \textit{simple} adjacency describes an upper bound on reachability, roughly similar to a Susceptible → Infected (SI) process, where there are no limits to how long an effect might remain in a vertex. The \textit{exponential}, and the discrete time variant \textit{geometric} temporal adjacency, roughly reminiscent of a Susceptible → Infected → Susceptible (SIS) process, allows vertices to remain affected for a duration of time determined through an exponential distribution with a given rate. The \textit{limited waiting-time} uses a deterministic maximum time cutoff instead of a probability distribution. Additionally, the \cpp library allows the definition of novel types of temporal network adjacency, which can then be used by the \cpp library function.

The future roadmap for this library focuses on the integration of the generalised compartmental model framework for static and temporal networks which was originally implemented for Ref.~\cite{rizi2022epidemic}, providing pre-compiled Python packages for x64 and ARMv8 macOS and Windows devices and implementing additional network statistics and algorithms.

\section{Illustrative Examples}
\subsection{Isotropic percolation in static networks}\label{sec:isotropic-example}
In this first example, we will focus on an analysis of static networks. This allows us to compare and contrast the interface and performance of the library with those of other network libraries. This example generates a set of $G(n, p)$ random networks and plots the largest connected component size as a function of $p$, highlighting the emergence of a giant component.

\begin{lstlisting}[language=Python]
import numpy as np
import reticula as ret
import matplotlib.pyplot as plt

from concurrent.futures import ThreadPoolExecutor
from functools import partial

def lcc_size(n, p):
    state = ret.mersenne_twister()
    g = ret.random_gnp_graph[ret.int64](n, p, state)
    return len(ret.largest_connected_component(g))

n = 1000000
ps = np.linspace(0, 2./n)

with ThreadPoolExecutor(max_workers=8) as e:
    lccs = e.map(partial(lcc_size, n), ps)

plt.plot(ps, list(lccs))
plt.show()
\end{lstlisting}

As discussed before, the public function interface of the library is very simple, as can be seen in the example above, for instance, in the call to the function \lstinline[language=Python]{ret.largest_connected_component}, which receives an undirected network as input and returns its largest connected component by number of vertices. The implementation of the $G(n, p)$ random graph model \lstinline[language=Python]{ret.random_gnp_graph[ret.int64]()}, on the other hand, requires information as to the data type of vertices which is provided in the square brackets. This serves a similar purpose to the \lstinline[language=Python]{dtype} parameter for numpy arrays: \lstinline[language=Python]{a = np.array([1, 2, 3], dtype=np.int64)}. Reticula, however, elected to use the more modern generic type interface that has been introduced to Python for use in type hints, where for example, \lstinline[language=Python]{list[int]} indicates a list of integers and \lstinline[language=Python]{dict[int, str]} indicates a dictionary with integer keys and string values \cite{rossum2014pep484}.

This allows for greater composability of types. For example, one can easily define a second order static undirected network type where each vertex is an undirected temporal edge with integers vertices and double precision floating point timestamps:
\begin{lstlisting}[language=Python]
g = ret.undirected_network[
      ret.undirected_temporal_edge[
        ret.int64, ret.double]]()
\end{lstlisting}

Use of the \lstinline[language=Python]{ThreadPoolExecutor} from the Python standard library demonstrates an example of using this library for parallel computation. The library itself makes no assumption about the computational environment, e.g., by naïvely trying to use as many CPUs as possible, as this approach might result in more problems than benefits in a high-performance computation environment. It is up to the user to specify the parallel computation model based on the set of tools provided by the programming language and other libraries specialising for this purpose. For example, in this case, the parameter \lstinline[language=Python]{maximum_workers} determines the maximum number 8 CPU cores are kept busy, which might have been determined based on memory limitations or the allocated number of CPUs. The library, in turn, provides safety guarantees for network and edge types and all the functions that consume these types and makes sure to release the global interpreter lock as soon as possible.

\begin{figure}
    \centering
    \includegraphics[width=\linewidth]{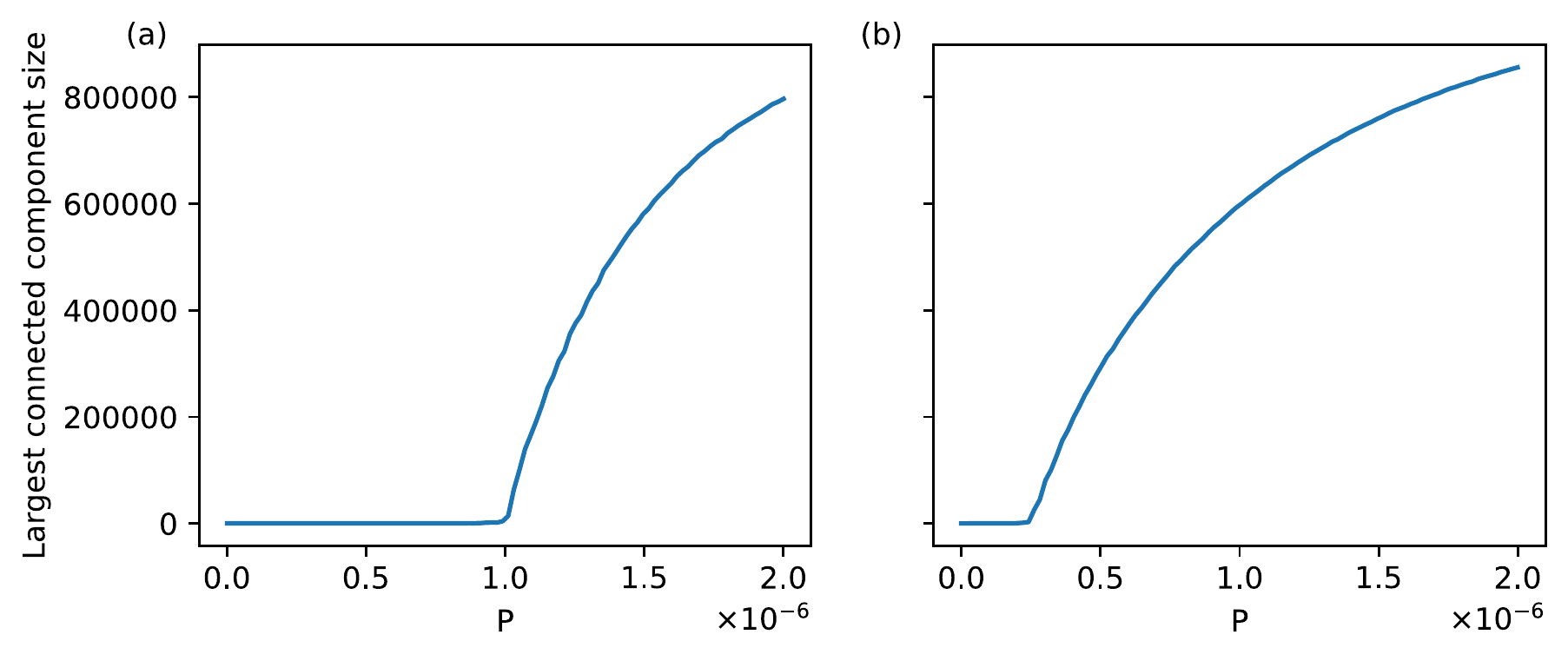}
    \caption{(a) Output figure of isotropic percolation example in Sec.~\ref{sec:isotropic-example} shows a phase transition in connectivity of random $G(n, p)$ static networks. (b) If instead of $G(n, p)$ networks, hypergraph networks with expected vertex degree $p (n-1)$ and edge degree 4 are generated, the phase transition shifts to smaller values of $p$.}
    \label{fig:isotropic-example}
\end{figure}

Running this example produces a plot similar to Fig.~\ref{fig:isotropic-example}(a), showing a clear phase transition at $p = p_c = 10^{-5}$, which is equivalent to the critical average degree and excess degree value $\langle k \rangle = 1$.

This example can be extended to random hypergraphs by generating e.g.~a random hypergraph with a given expected degree-sequence instead of a random (dyadic) $G(n, p)$ network. This can be accomplished by swapping the use of the function \lstinline[language=Python]{ret.random_gnp_graph} with
\begin{lstlisting}[language=Python]
g = ret.random_expected_degree_sequence_hypergraph[ret.int64](
        vertex_weight_sequence=[p*(n-1)]*n,
        edge_weight_sequence=[4]*int((p*(n-1)*n)/4),
        random_state=state)
\end{lstlisting}
which generates a static hypergraph with expected vertex degree of $p (n-1)$ and edge degree 4 for large values of $n$ using a generalisation of the Chung-Lu algorithm to bipartite networks \cite{aksoy2017measuring}. This creates a plot with the critical threshold at $2.5\times10^{-6}$, shown in Fig.~\ref{fig:isotropic-example}(b).

\subsection{Temporal network event time inhomogeneity and reachability}
For this example, we generate a temporal network from a random $G(n, p)$ static network and self-exciting process activation times, shuffle away the correlation of the inter-event times using one of the implemented microcanonical randomised reference models and compare the change in reachability.

\begin{lstlisting}[language=Python]
import reticula as ret
import reticula.microcanonical_reference_models as mrrm

from concurrent.futures import ThreadPoolExecutor
from functools import partial

def out_cluster_mass_at_t0(network, vertex):
    return ret.out_cluster(
            temporal_network=network, temporal_adjacency=adj,
            vertex=vertex, time=0.0).mass()

n = 128
max_t = 256
state = ret.mersenne_twister()
ens = 100

self_exceiting_mean_masses = []
randomised_mean_masses = []
with ThreadPoolExecutor(max_workers=8) as e:
    for _ in range(ens):
        base_net = ret.random_gnp_graph[ret.int64](
                n=n, p=2/n, random_state=state)
        iet_dist = ret.hawkes_univariate_exponential[ret.double](
                mu=0.2, alpha=0.8, theta=0.5)
        g = ret.random_link_activation_temporal_network(
                base_net, max_t, iet_dist, state)
        adj = ret.temporal_adjacency.exponential[g.edge_type()](
                rate=1.0, seed=0)
        cluster_masses_g = list(e.map(
            partial(out_cluster_mass_at_t0, g),
            g.vertices()))
        self_exceiting_mean_masses.append(
                sum(cluster_masses_g)/len(cluster_masses_g))

        randomised = mrrm.timeline_shuffling(
                temporal_network=g, random_state=state,
                t_start=0, t_end=max_t)
        cluster_masses_rand = list(e.map(
            partial(out_cluster_mass_at_t0, randomised),
            randomised.vertices()))
        randomised_mean_masses.append(
                sum(cluster_masses_rand)/len(cluster_masses_rand))

print("mean mass with self-exciting inter-event times:",
        sum(self_exceiting_mean_masses)/len(self_exceiting_mean_masses))
print("mean mass after timeline shuffling randomisation:",
        sum(randomised_mean_masses)/len(self_exceiting_mean_masses))
\end{lstlisting}

This example produces \lstinline[language=Python]{ens = 100} random temporal networks from a random $G(n, p)$ static network and the Hawkes univariate exponential self-exciting activation times. For each random network, it calculates the mean temporal out-cluster mass. The temporal network is then randomised to remove the correlation between consecutive event times (the self-excitement) using the timeline shuffling method \cite{gauvin2022randomized} and mean temporal out-cluster masses are calculated again. In this example, we used the exponential model of temporal adjacency, meaning that the component mass conceptually resembles the total human-hours of infection in an SIS model where the transition from Infected back to Susceptible is governed by an exponential time with a rate parameter, which set to 1.0 in this case.

The mean temporal out-cluster mass is calculated by computing out-clusters from every vertex at $t=0$ (in function \lstinline[language=Python]{out_cluster_mass_at_t0}) and averaging over the component mass. It would be possible to use the much faster and have a larger sample size of starting points and times by using a probabilistic counting method to estimate all possible out-clusters, but care has to be given to the fact that the event times in this example are not distributed homogeneously.

Running the example produces a result similar to below
\begin{lstlisting}
mean mass with self-exciting inter-event times: 2164.3724129038064
mean mass after timeline shuffling randomisation: 5539.766618496214
\end{lstlisting}
which can be interpreted as follows: Shuffling away the self-exciting property of a random temporal network generated with the parameters presented substantially increases the average reachability of the temporal network. Although a proper study of such hypothesis requires stronger analytical methods, e.g.~in establishing significance of difference in means masses, or to confirm this on empirical networks and a wider set of synthetic networks, all of which is outside the scope of this example.

Note that the temporal networks are not copied to each thread, but a single instance of the temporal network lives in the memory at each point in time. Using network types in shared memory across threads is safe in Reticula, as it is not possible to, perhaps inadvertently, modify the network from one thread while reading from another thread. In this example, the user can be sure, without having to check the documentation or the implementation, that the function \lstinline[language=Python]{ret.out_cluster} or any other function that takes a network as an input does not modify the argument. This has the effect of reducing the risk of inadvertent data races and is a result of the fact that the network types in Reticula are immutable. This is not guaranteed for other libraries presented in Tab.~\ref{tab:compare}, which makes it hard to reason about possible data races for igraph and SNAP network types, two network libraries implemented as native extensions which also support multi-threaded processing.

Similar to the previous example in Sec.~\ref{sec:isotropic-example}, this example can also be extended seamlessly to generate hypergraph temporal networks, for example by replacing the use of \lstinline[language=Python]{ret.random_gnp_graph} with
\begin{lstlisting}[language=Python]
base_net = ret.random_expected_degree_sequence_hypergraph[ret.int64](
        vertex_weight_sequence=[2]*n,
        edge_weight_sequence=[4]*n//2,
        random_state=state)
\end{lstlisting}
which results in the following function \lstinline[language=Python]{ret.random_link_activation_temporal_network} generating a hypergraph temporal network instead of a dyadic temporal network. This illustrates how the same API---in the case of these illustrative examples generating an edge activation network or calculating temporal clusters or largest connected components of static networks---can operate on different network types to the extent that those operations are well-defined for the given type of network.

\section{Conclusion}

Reticula is a software library that natively handles directed and undirected static networks, temporal networks, hypergraphs and temporal hypergraphs with a unified programming interface with either Python or \cpp. The library optimises the use of CPU cache and provides thread-safe types and operations, making it suitable to use in a multi-threaded settings in high-performance computation environment as well as on modern, high core-count CPUs. Reticula enables the scientists to study the properties of very large network datasets, construct and compare reference models and random networks, and read and write networks to the disk.

\section*{Acknowledgements}
\label{sec:acknowledgements}

We acknowledge the computational resources and technical consultation provided by the Aalto Science-IT project.

\bibliographystyle{elsarticle-num} 
\bibliography{references}
\end{document}